\documentclass[9pt,conference]{IEEEtran}
\IEEEoverridecommandlockouts

\usepackage{cite}
\usepackage{amsmath,amssymb,amsfonts}
\usepackage{algorithmic}
\usepackage{graphicx}
\usepackage{textcomp}
\usepackage{xcolor}
\usepackage{microtype}
\usepackage{hyperref}
\usepackage{url}
\usepackage{booktabs}
\usepackage{textcomp}
\usepackage{multirow}
\usepackage{multicol}
\usepackage{makecell}
\usepackage{subfigure}
\usepackage{latexsym}
\usepackage{enumitem}
\usepackage{subcaption}
\usepackage{wrapfig}
\usepackage{caption}
\mathchardef\mhyphen="2D

\def\BibTeX{{\rm B\kern-.05em{\sc i\kern-.025em b}\kern-.08em
    T\kern-.1667em\lower.7ex\hbox{E}\kern-.125emX}}
\begin{document}

\title{BigCodec: Pushing the Limits of Low-Bitrate\\Neural Speech Codec
\thanks{This work was supported by JSPS KAKENHI, Grant Number JP23KJ0828 and 22H03639, and JST FOREST JPMJFR226V.}}

\author{
\begin{tabular}{c}
Detai Xin$^{1}$, Xu Tan$^{2}$, Shinnosuke Takamichi$^{3}$, Hiroshi Saruwatari$^{1}$\end{tabular}\\
$^{1}$The University of Tokyo, $^{2}$Microsoft, $^{3}$Keio University\\
\href{mailto:detai\_xin@ipc.i.u-tokyo.ac.jp}{\texttt{detai\_xin@ipc.i.u-tokyo.ac.jp}}
}

\maketitle
\setlength\textfloatsep{5mm} 
\setlength{\abovedisplayskip}{3pt} 
\setlength{\belowdisplayskip}{3pt} 
\setlength{\tabcolsep}{0.9mm} 
\allowdisplaybreaks

\begin{abstract}
We present BigCodec, a low-bitrate neural speech codec.
While recent neural speech codecs have shown impressive progress, their performance significantly deteriorates at low bitrates (around 1 kbps).
Although a low bitrate inherently restricts performance, other factors, such as model capacity, also hinder further improvements.
To address this problem, we scale up the model size to 159M parameters that is more than 10 times larger than popular codecs with about 10M parameters.
Besides, we integrate sequential models into traditional convolutional architectures to better capture temporal dependency and adopt low-dimensional vector quantization to ensure a high code utilization.
Comprehensive objective and subjective evaluations show that BigCodec, with a bitrate of 1.04 kbps, significantly outperforms several existing low-bitrate codecs.
Furthermore, BigCodec achieves objective performance comparable to popular codecs operating at 4-6 times higher bitrates, and even delivers better subjective perceptual quality than the ground truth.
\end{abstract}


\vspace{-3mm}
\section{Introduction}
\vspace{-1mm}
\label{section:introduction}
Neural speech/audio codecs have made rapid advancement in recent years.
The primary goal of a speech codec is to compress audio signals into lossy discrete representations at lower bitrates while preserving perceptual quality as much as possible.
Speech codecs play an important role in data transmission as they help reduce the network traffic.
While traditional speech codecs~\cite{opus2012, evs2015} construct complex pipelines, recent speech codecs extensively adopt end-to-end deep neural networks (DNNs) with vector quantization (VQ) to build the model~\cite{zeghidour2021soundstream, jayashankar2022architecture, defossez2022encodec, kumar2024descript}.
Such DNN-based codecs show better reconstruction fidelity than previous works.
And the bitrates typically range from $6$~kbps to $12$~kbps for $16$ kHz audio data.

Despite the impressive progress, the theoretical bitrate of speech appears to be much lower than that of current codecs.
Denes et al.~\cite{denes1963statistics} estimate the information rate of English based on phoneme statistics and find the lexical bitrate is only about $50$~bps.
Van et al.~\cite{van2017information} use the information bottleneck principle~\cite{tishby2000bottleneck} to estimate the information rate of speech, resulting in an upper bound of $100$~bps that is merely twice the lexical bitrate.
These theoretical bitrates suggest that the bitrates of current codecs can be potentially reduced.
Achieving such a reduction could further enhance communication efficiency by enabling high-quality audio transmission with minimal data usage.

However, in practice, when the bitrate of a codec drops to about $1$~kbps, the reconstruction performance degrades significantly~\cite{defossez2022encodec}.
Previous works approach this problem through various aspects, including neural architectures~\cite{kleijn2018wavenet, siahkoohi2022ultra, zheng2024srcodec, zheng2024supercodec}, coding methods~\cite{jiang2023latent}, and non-deterministic encoding~\cite{jenrungrot2023lmcodec}.
But none of these studies explored the potential of scaling up model size.
Increasing model size has proven effective in several domains, such as language generation~\cite{brown2020gpt3} and image synthesis~\cite{ramesh2021text2image}.
In audio/speech generation, BigVGAN~\cite{lee2022bigvgan}, as one of the most prominent examples of scaling up model size in this domain, also demonstrates its effectiveness.

In this paper, we introduce a neural speech codec with a bitrate of $1.04$~kbps and $159$M parameters.
We name this codec BigCodec, as our idea of scale-up is similar to BigVGAN~\cite{lee2022bigvgan}.
We adopt the popular generative adversarial network (GAN) framework~\cite{goodfellow2020generative}, and incorporate techniques from previous state-of-the-art (SOTA) codecs~\cite{kumar2024descript} to ensure a high codebook utilization.
BigCodec features two key distinctions from prior work.
First, BigCodec has $159$M parameters that is about $11$ times larger than previous typical codecs with about $14$M parameters~\cite{defossez2022encodec}.
Second, BigCodec uses a single codebook with $8192$ codes and vanilla VQ-VAE~\cite{van2017vqvae} for quantization, which is different from most previous low-bitrate codecs that use multiple codebooks~\cite{siahkoohi2022ultra, jenrungrot2023lmcodec, jiang2023latent, zheng2024supercodec, zheng2024srcodec} or sequential quantization method~\cite{jiang2023latent} that brings extra complexity.
We train BigCodec on the LibriSpeech~\cite{panayotov2015librispeech} corpus with $960$ hours of speech data.
Comprehensive objective and subjective evaluations demonstrate that BigCodec achieves significantly better performance than previous low-bitrate codecs.
Furthermore, BigCodec demonstrates comparable objective performance to popular neural speech codecs that have $6$ times (Encodec~\cite{defossez2022encodec}) or $4$ times (descript speech codec (DAC)~\cite{kumar2024descript}) higher bitrates, and even delivers better subjective perceptual quality than the ground truth (GT).
We conduct ablation studies to highlight the contributions of scale-up and our design choices.
Besides, we evaluate BigCodec from various aspects including the performance on unseen languages, code utilization, and inference efficiency.
The contributions are summarized as follows:
\begin{itemize}[leftmargin=*]
    \item We introduce BigCodec, a low-bitrate neural speech codec with a bitrate of $1.04$~kbps. The model is scaled up to $159$M parameters, and it employs a single codebook with $8192$ codes that avoids extra complexity of multiple codebooks.
    \item We conduct comprehensive objective and subjective evaluations to demonstrate that BigCodec significantly outperforms previous low-bitrate codecs, and shows comparable objective performance to previous codecs operating at 4-6 times higher bitrates. BigCodec even delivers better subjective perceptual quality than GT samples.
    \item We conduct ablation studies and evaluate BigCodec's generalization ability and inference efficiency.
\end{itemize}
Audio samples can be found at \url{https://aria-k-alethia.github.io/bigcodec-demo/}.
We publicate inference scripts and pretrained checkpoints of BigCodec at \url{https://github.com/Aria-K-Alethia/BigCodec}.
\vspace{-2mm}
\section{Related work}
\vspace{-1mm}
Several previous works attempt codecs with low bitrates around $1$~kbps.
Due to the inherent constraints of low bitrates, many of these methods restrict their training data to specific domains, such as speech~\cite{jiang2023latent, jenrungrot2023lmcodec, zheng2024supercodec, zheng2024srcodec}, and improve the performance using various methods.
One important method is enhancing encoder/decoder.
Kleijn et al.~\cite{kleijn2018wavenet} propose to leverage a powerful audio generation model WaveNet for decoding.
Jenrungrot et al.~\cite{jenrungrot2023lmcodec} propose a non-deterministic encoding method that only transmits codes of lower layers of residual VQ (RVQ)~\cite{zeghidour2021soundstream} and uses a Transformer to predict codes of higher layers.
Another promising direction is modeling temporal dependency to better capture speech structure.
Siahkoohi et al.~\cite{siahkoohi2022ultra} combine a convolutional neural network (CNN) with a pretrained Transformer, using the latter to model long-distance dependency.
Jiang et al.~\cite{jiang2023latent} propose a novel method named TF-Codec, which employs predictive coding~\cite{atal1982predictive} to autoregressively encode temporal residuals, thereby minimizing redundancy over time compared to independent frame processing.
Lastly, some works explore domain-specific optimizations.
Yang et al.~\cite{yang2024uniaudio15} propose a speech codec named LLM-Codec that initializes the codebook from a pretrained large language model to adapt the model to fit speech data.
\vspace{-3mm}
\section{BigCodec}
\vspace{-1mm}
\label{section:proposed}
We adopt the prevalent GAN framework~\cite{goodfellow2020generative} comprising a VQ-VAE generator and multiple discriminators.
In the subsequent sections, we first detail the architecture, then outline the objective for training, and finally explain how we scale up the model size.

\begin{figure}[t]
  \centering
  \includegraphics[width=\linewidth]{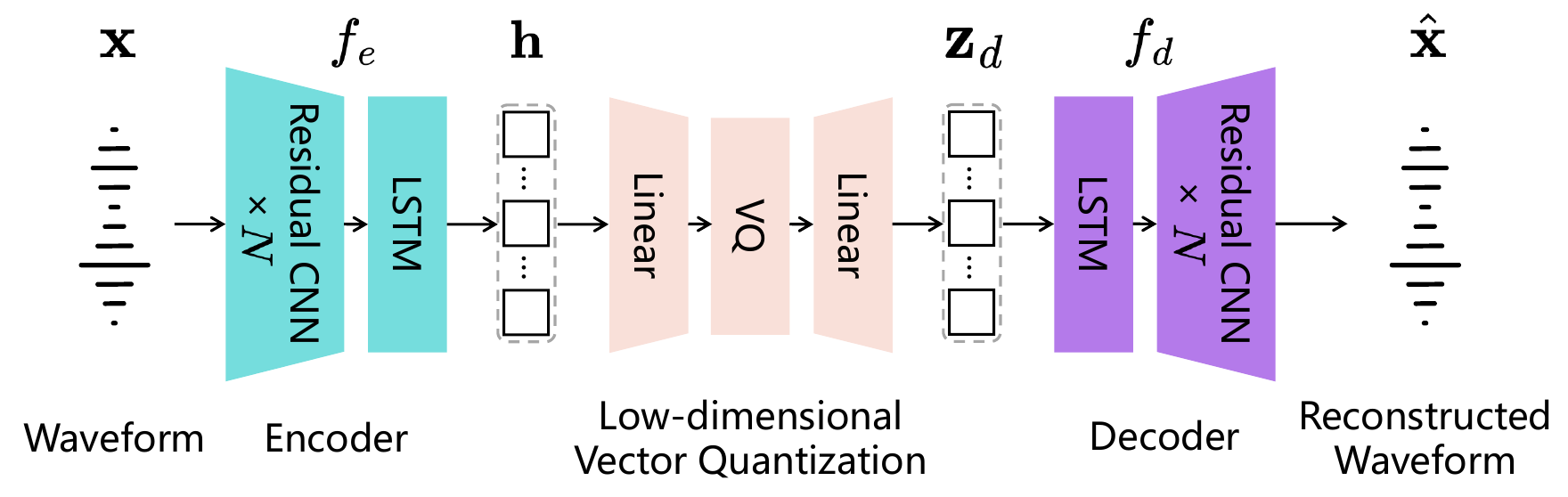}
  \caption{Architecture of the VQ-VAE generator of BigCodec. All symbols are defined in Section~\ref{subsection:architecture}.}
  \label{fig:generator}
  \vspace{-3mm}
\end{figure}

\vspace{-2mm}
\subsection{Architecture}
\vspace{-1mm}
\label{subsection:architecture}
\subsubsection{VQ-VAE generator}
\paragraph{Encoder \& Decoder}
The general architecture of the VQ-VAE generator in BigCodec is illustrated in Fig~\ref{fig:generator}.
The encoder and decoder have mirror structures, hence their computation flows are reversed.
The encoder consists of several residual CNN blocks~\cite{kumar2024descript} with snake activation functions~\cite{ziyin2020snake} and a two-layer unidirectional long short-term memory (LSTM) network.
Each CNN block downsamples the waveforms by a factor, and has multiple convolutions with different dilation rates to recognize patterns in sequential data.
The LSTM is used to model long-distance dependency, and we found that it consistently improved the performance in the preliminary experiments.

Formally, denote the input waveform, the encoder, and the output hidden variables as $\mathbf{x}=\{x_{1}, x_{2}, \ldots, x_{T_{x}}\}$, $f_{e}(\cdot)$, and $\mathbf{h}=\{h_{1}, h_{2}, \ldots, h_{T_{h}}\}$, respectively.
Then, we have $\mathbf{h} = f_{e}(\mathbf{x})$.
Assume there are $N$ CNN blocks in the encoder and the downsampling factor of the $i$-th block is $r_{i}$, the total downsampling rate of the encoder is: $R=\prod_{i=1}^{N} r_{i}$, which leads to $T_{x} = R \times T_{h}$.

The decoder $f_{d}(\cdot)$ takes the output $\mathbf{z}_{d}$ of the VQ module (described in the next section) as input and outputs the reconstructed waveform $\hat{\mathbf{x}}$, i.e., $\hat{\mathbf{x}}=f_{d}(\mathbf{z}_{d})$.
Each CNN block in the decoder upsamples the input using transposed convolutions with the same rates to the encoder.

\paragraph{Vector quantization}
The vector quantization module quantizes $\mathbf{h}$ into vectors from a fixed-size codebook.
We use a single codebook to limit the bitrate and simplify the quantization process, so each frame is represented by a single code.
The low bitrate makes it crucial to ensure a high codebook utilization, hence we adopt a technique proposed in Yu et al.~\cite{yu2021vector} that projects the latent variables into a low-dimensional space before quantization.
The quantization is performed by finding the vector in the codebook with the smallest L2 distance to the latent variables.
To further enhance performance, we L2-normalize both the latent variables and the codebook vectors, as suggested in Yu et al.~\cite{yu2021vector}, which makes the L2 distance become cosine distance.
The quantized vectors are then projected back to the original dimension of $\mathbf{h}$, resulting in $\mathbf{z}_{d}$ that serves as the input to the decoder described in the previous section.
We find that projecting $\mathbf{h}$ into a low-dimensional space substantially improves the codebook utilization, as it avoids to quantize in the sparse high-dimensional space of $\mathbf{h}$.

\subsubsection{Discriminators}
The discriminators are trained to differentiate between real and generated data.
We identify three main types of discriminators that have proven particularly effective for speech/audio:
(1) The multi-scale discriminator (MSD) proposed in MelGAN~\cite{kumar2019melgan}, which detects consecutive temporal patterns in waveforms at various scales.
(2) The multi-period discriminator (MPD) introduced in HiFi-GAN~\cite{kong2020hifi}, which captures multiple periodic patterns in waveforms.
(3) The multi-scale short-time Fourier transform (MS-STFT) discriminator used in EnCodec~\cite{defossez2022encodec} and the multi-resolution discriminator (MRD) from UnivNet~\cite{jang21univnet}, both of which learn spectral structures in spectrograms at multiple resolutions.
In BigCodec, we adopt the MPD and MS-STFT discriminators.
Our preliminary experiments showed that these two discriminators were effective, and using the MSD did not provide additional benefits, which is also observed in EnCodec~\cite{defossez2022encodec}.

\begin{table*}[t]
    \centering
    \caption{Main results of BigCodec on the LibriSpeech test set with $2620$ utterances. \textbf{Bold} indicates the best score with $p < 1e\mhyphen3$ compared to previous low-bitrate codecs. For MUSHRA both the mean scores and the $95\%$ confidence intervals are reported. The inter-rater agreement of the MUSHRA test is revealed by an ICC(2) of $0.73$, as described in Section~\ref{subsection:main_results}.}
    \begin{tabular}{l|ccc|ccccc|c}
    \toprule
     & \#Params. (M) & \makecell[c]{Bitrate\\(kbps)}&  \makecell[c]{Approx. bitrate\\(kbps)} & MCD $\downarrow$  & PESQ-NB $\uparrow$ & PESQ-WB $\uparrow$ & STOI $\uparrow$ & SIM $\uparrow$ & MUSHRA $\uparrow$ \\
    \midrule
    GT & - &$16$ & - &$0.00$ & $4.54$ & $4.64$ & $1.00$ & $1.00$ & $91.84 {\scriptstyle \pm 1.31}$\\
    DAC-4k~\cite{kumar2024descript} & $74$ &$4$ & $3.88$ &$4.27$ & $3.19$ & $2.72$ & $0.94$ & $0.87$ & $76.46 {\scriptstyle \pm 2.23}$\\
    EnCodec-6k~\cite{defossez2022encodec} & $14$ & $6$ & $4.83$ & $4.14$ & $3.18$ & $2.77$ & $0.94$ & $0.89$ & $70.11 {\scriptstyle \pm 2.32}$ \\
    \midrule
    DAC-1k~\cite{kumar2024descript} & $74$ &$1$ & $0.92$ & $8.97$ & $1.40$ & $1.13$ & $0.73$ & $0.32$ & $11.91 {\scriptstyle \pm 1.28}$ \\
    EnCodec-1.5k~\cite{defossez2022encodec} & $14$ & $1.5$ & $1.22$ & $6.16$ & $1.94$ & $1.56$ & $0.85$ & $0.60$ & $29.94 {\scriptstyle \pm 1.97}$\\
    LLM-Codec~\cite{yang2024uniaudio15} & $38$ & $0.74$ & $0.68$ &$6.08$ & $2.49$ & $1.97$ & $0.88$ & $0.64$ & $67.90 {\scriptstyle \pm 2.46}$ \\
    TF-Codec~\cite{jiang2023latent} & $6.37$ & $1.5$ & $1.21$ & $5.16$ & $3.11$ & $2.53$ & $0.92$ & $0.73$ & $74.67 {\scriptstyle \pm 2.36}$\\
    \midrule
    BigCodec & $159$ & $1.04$ & $1.03$ &$\mathbf{4.55}$ & $\mathbf{3.27}$ & $\mathbf{2.68}$ & $\mathbf{0.93}$ & $\mathbf{0.84}$ & $\mathbf{92.33} {\scriptstyle \pm 1.29}$\\
    \bottomrule
    \end{tabular}
    \label{tab:main_results}
\end{table*}
\vspace{-2mm}
\subsection{Training objective}
\vspace{-1mm}
\paragraph{Reconstruction loss}
We use the multi-scale mel-spectogram reconstruction loss proposed in DAC~\cite{kumar2024descript} that measures the L1 distance in the spectral domain at multiple scales.
We found this loss to be more effective than the multi-scale STFT loss~\cite{yamamoto2020parallel, zeghidour2021soundstream} because the mel-spectrogram directly relate to perceptual quality.
The time-domain reconstruction L1 loss~\cite{defossez2022encodec} is not used as we found it produced blurry results.

\paragraph{GAN loss}
We use the loss function of least-square GAN~\cite{mao2017lsgan} to train the model, which uses least square loss instead of conventional binary cross entropy loss to stabilize training.
Additionally, the L1 feature matching loss proposed in HiFi-GAN~\cite{kong2020hifi} is incorporated.

\paragraph{VQ loss}
The codebook is directly learned by the L1 loss between the projected encoder output and the quantized result with a stop-gradient operator applied~\cite{van2017vqvae}.
We do not use moving average to update the codebook~\cite{van2017vqvae, zeghidour2021soundstream, defossez2022encodec} as we found that the codebook could be effectively learned by gradient descent in the low-dimensional space.
A commitment loss is added to prevent the output of encoder from growing excessively~\cite{van2017vqvae}.
Since the VQ module contains a non-differentiable \texttt{argmin} operation, we employ the straight-through estimator~\cite{bengio2013straight_through} to back-propagate the gradient.

\paragraph{Weighting}
The final training objective of BigCodec is the weighted sum of all aforementioned losses.
The weight for the multi-scale mel-spectrogram reconstruction loss is set to $15$, as it directly relates to speech quality.
The commitment loss in the VQ module is weighted at $0.25$ to prevent mode collapse.
All other losses are assigned a weight of $1$.

\vspace{-2mm}
\subsection{Scaling up the model size}
\vspace{-1mm}
We adopt a similar strategy to BigVGAN~\cite{lee2022bigvgan} for scaling up the model.
First, a basic version, named BigCodec-base, is constructed with about $17$M parameters (excluding the parameters of the discriminators).
The number of CNN blocks in the encoder/decoder, i.e., $N$, is set to $4$.
The channel size of the input starts at $32$ and doubles with each CNN block, resulting in a $512$-dimensional $\mathbf{h}$.
Note again that the decoder mirrors the structure of the encoder.

Next, we scale up the model size.
We increase $N$ to $5$ while keeping the total downsampling rate $R$ fixed.
Accordingly, the channel size is increased, starting at $48$ and reaching $1536$ after the final CNN block in the encoder.
The final version of BigCodec contains $159$M parameters in the VQ-VAE and $22$M parameters in the discriminators.
\vspace{-3mm}
\section{Experiments}
\vspace{-2mm}
\subsection{Setup}
\label{subsection:setup}
We use the full training set of LibriSpeech~\cite{panayotov2015librispeech} with $960$ hours of speech data for training.
The \texttt{test-clean} set with $2620$ utterances is used for testing.
All speech data are in $16$~kHz.

The total downsampling rate $R$ is set to $200$, and the codebook size is set to $8192$, resulting in a theoretical bitrate of $16000 / R \times \log{8192}=1.04$~kbps for BigCodec.
The projection dimension in the VQ module is set to $8$, as recommended in DAC~\cite{kumar2024descript}.

All models are trained on 8 NVIDIA A100 GPUs.
The batch size is set to $8$, and each item is a $1$-second segment randomly cropped from the original utterance.
We use AdamW~\cite{loshchilov2018adamw} as the optimizer, with the moving average coefficients $\beta_{1}$ and $\beta_{2}$ set to $0.8$ and $0.9$, respectively.
A scheduled learning rate that linearly declines from $1e\mhyphen4$ to $1e\mhyphen5$ with $1$k warmup steps is used.
The model converges in about $600$k steps.

We compare BigCodec with the following popular codecs and previous low-bitrate codecs:
\begin{itemize}[leftmargin=*]
    \item \textbf{EnCodec}~\cite{defossez2022encodec}: A popular neural codec using RVQ. The model has $14$M parameters and is trained on a large and diverse dataset with about $17$k hours of multilingual speech, music, and sound data. We use the official $24$~kHz checkpoint\footnote{\url{https://github.com/facebookresearch/encodec}} and downsample the utterances to $16$~kHz after inference. Since there are multiple available bitrates of EnCodec, we evaluate it at two bitrates: (1) EnCodec-1.5k, which is closest to $1.04$~kbps, and (2) EnCodec-6k, which serves as an upper bound with performance comparable to BigCodec.
    \item \textbf{DAC}~\cite{kumar2024descript}: A SOTA codec that combines RVQ with the low-dimensional VQ technique. The model has $74$M parameters and is also trained on a large and diverse dataset like EnCodec. We use the official $16$~kHz checkpoint\footnote{\url{https://github.com/descriptinc/descript-audio-codec}}. As with EnCodec, we evaluate DAC at two bitrates: DAC-1k and DAC-4k.
    \item \textbf{TF-Codec}~\cite{jiang2023latent}: A SOTA low-bitrate speech codec utilizing predictive coding. The model has $6.37$M parameters and is trained on $890$ hours of clean speech from the Deep Noise Suppression Challenge at ICASSP 2021~\cite{reddy2021icassp}, which includes multilingual, emotional, and singing data. The results are obtained from the authors.
    \item \textbf{LLM-Codec}~\cite{yang2024uniaudio15}: A recent low-bitrate codec specifically designed for speech. The model has $38$M parameters and is trained on $2$k hours of data from the multilingual LibriSpeech (MLS) corpus~\cite{pratap20mls} and the AudioCaps sound corpus~\cite{kim2019audiocaps}. We use the official checkpoint\footnote{\url{https://github.com/yangdongchao/LLM-Codec}}.
\end{itemize}
Besides, we also evaluated SRCodec~\cite{zheng2024srcodec}, SuperCodec~\cite{zheng2024supercodec}, and LMCodec~\cite{jenrungrot2023lmcodec}.
However, these models were either found to have relatively weak performance or, in the case of LMCodec, only had a small number of available test samples.
Therefore, we omit the comparisons with these methods.

We adopt the following objective metrics to evaluate performance:
\begin{itemize}[leftmargin=*]
    \item \textbf{Approximated bitrate (Approx. bitrate)}: Computed by approximating the information entropy on the test set. An approximated bitrate close to the theoretical bitrate indicates a high codebook utilization.
    \item \textbf{Mel cepstral distortion (MCD)}: Directly measures reconstruction quality in the spectral domain.
    \item \textbf{PESQ}~\cite{rix2001pesq}: An intrusive metric that assesses the perceptual quality of speech. We report both narrow band (PESQ-NB) and wide band (PESQ-WB) results.
    \item \textbf{STOI}~\cite{taal2011stoi}: An intrusive metric that evaluates speech intelligibility. We found STOI to be strongly related to perceptual intelligibility.
    \item \textbf{Speaker similarity (SIM)}: Defined as the cosine similarity between the speaker embeddings of the GT and the reconstructed utterances. Although this metric is rarely used in previous works, we argue that it is crucial for low-bitrate codecs, as the low bitrate restricts the model's ability to encode detailed acoustic information such as timbre. As one will see in Section~\ref{subsection:main_results}, with a low bitrate the SIM metric of some codecs drops significantly even if they have fairly good intelligibility.   We use the \texttt{
wavlm\_large\_finetune} checkpoint of WavLM-TDNN\footnote{\url{https://github.com/microsoft/UniSpeech/tree/main/downstreams/speaker_verification}}, a speaker verification model based on WavLM~\cite{chen2022wavlm}, to extract speaker embeddings.
\end{itemize}
To evaluate the subjective perceptual quality, we adopt the MUSHRA protocol~\cite{series2014mushra} with a hidden reference as the anchor.
We do not use the low-passed anchor.
The evaluation interface is based on webMUSHRA~\cite{schoeffler2018webmushra}.
We randomly select $10$ utterances from the test set and recruit $30$ native speakers on Prolific\footnote{\url{https://www.prolific.com/}} to complete the test.
Each participant is compensated \pounds $2.25$ based on an hourly salary of \pounds $9$.
We exclude results from participants whose mean score for the reference is less than $80$, resulting in $29$ valid participants.
Each model finally receives $290$ scores.

\vspace{-2mm}
\subsection{Main results}
\vspace{-1mm}
\label{subsection:main_results}
We first evaluate BigCodec and other baselines on the \texttt{test-clean} set of LibriSpeech with $2620$ utterances.
The results are shown in Table~\ref{tab:main_results}.
First, it can be seen that BigCodec consistently outperforms previous low-bitrate codecs across all objective metrics, demonstrating its superior ability to reconstruct speech with better perceptual quality, intelligibility, and speaker similarity.
Second, BigCodec achieves performance comparable to that of higher-bitrate codecs, i.e., DAC-4k and EnCodec-6k, further highlighting the effectiveness of scaling up the model size in BigCodec.
Notably, BigCodec shows a significant improvement in speaker similarity (SIM) over previous works, even in cases where codecs like TF-Codec achieve similar intelligibility (STOI).
This indicates that BigCodec excels not only in reconstructing content information but also in preserving acoustic details.
Third, BigCodec obtains the highest MUSHRA score of $92.33$, which surprisingly outperforms all baselines by a large margin.
Remarkably, BigCodec's MUSHRA score is even slightly higher than that of the GT ($91.84$), demonstrating the superior perceptual quality achieved by BigCodec.
We strongly encourage readers to listen to the audio samples to appreciate the perceptual quality firsthand.
The inter-rater agreement is assessed by the intraclass correlation coefficient (ICC).
For this test, we compute ICC(2) based on the two-way random effects model, resulting in a value of $0.73$, which indicates a good agreement as suggested by Cicchetti et al.~\cite{cicchetti1994guidelines}.
Finally, the approximated bitrate of BigCodec ($1.03$ kbps) is almost identical to the theoretical bitrate ($1.04$ kbps), indicating the effectiveness of the low-dimensional vector quantization.

\begin{table}[t]
    \centering
    \caption{Multilingual evaluation results of BigCodec on the MLS test set with $700$ utterances from $7$ OOD languages. Note that BigCodec is the \emph{only} codec that is trained on a monolingual corpus (English). \textbf{Bold} indicates the best score with $p < 1e\mhyphen3$ compared to previous low-bitrate codecs.}
    \footnotesize
    \begin{tabular}{l|ccccc}
    \toprule
     & MCD $\downarrow$  & PESQ-NB $\uparrow$ & PESQ-WB $\uparrow$ & STOI $\uparrow$ & SIM $\uparrow$ \\
    \midrule
    DAC-4k~\cite{kumar2024descript} & $4.37$ & $3.13$ & $2.59$ & $0.93$ & $0.91$ \\
    EnCodec-6k~\cite{defossez2022encodec} & $4.12$ & $3.19$ & $2.72$ & $0.94$ & $0.92$ \\
    \midrule
    DAC-1k~\cite{kumar2024descript} & $8.85$ & $1.40$ & $1.12$ & $0.73$ & $0.33$ \\
    EnCodec-1.5k~\cite{defossez2022encodec} & $6.07$ & $1.99$ & $1.56$ & $0.84$ & $0.65$ \\
    LLM-Codec~\cite{yang2024uniaudio15} & $6.15$ & $2.39$ & $1.87$ & $0.87$ & $0.69$ \\
    TF-Codec~\cite{jiang2023latent} & $5.26$ & $3.00$ & $2.41$ & $0.91$ & $0.78$ \\
    \midrule
    BigCodec & $\mathbf{4.86}$ & $\mathbf{3.09}$ & $\mathbf{2.47}$ & $\mathbf{0.92}$ & $\mathbf{0.86}$ \\
    \bottomrule
    \end{tabular}
    \label{tab:mls_results}
\end{table}
\vspace{-2mm}
\subsection{Unseen languages}
\vspace{-1mm}
\label{subsection:ood_results}
Second, we assess the generalization ability of BigCodec on unseen languages.
With a limited bitrate, achieving robust generalization becomes more challenging for the codecs.
We use the MLS~\cite{pratap20mls} corpus and randomly select $100$ utterances for each of the $7$ languages (German, Dutch, Spanish, French, Italian, Portuguese, Polish) except for English, forming a test set with $700$ utterances.

The results are shown in Table~\ref{tab:mls_results}.
It is noteworthy that BigCodec is the only codec in the table trained on a monolingual corpus.
Although the performance of BigCodec decreases to some extent compared to the in-domain results in Table~\ref{tab:main_results}, it still significantly outperforms all previous low-bitrate codecs, demonstrating its strong generalization ability across unseen languages. 
Besides, we notice an increase in the SIM scores for almost all codecs, which we attribute to the relatively small number of speakers in this test set.

\begin{table}[t]
    \centering
    \caption{Results of ablation studies on the LibriSpeech test set with $2620$ utterances. \textbf{Bold} indicates the best scores with $p < 1e\mhyphen3$.}
    \footnotesize
    \begin{tabular}{l|ccccc}
    \toprule
     & MCD $\downarrow$  & PESQ-NB $\uparrow$ & PESQ-WB $\uparrow$ & STOI $\uparrow$ & SIM $\uparrow$ \\
    \midrule
    BigCodec & $\mathbf{4.55}$ & $\mathbf{3.27}$ & $\mathbf{2.68}$ & $\mathbf{0.93}$ & $\mathbf{0.84}$ \\
    \midrule
    BigCodec-base & $4.87$ & $3.03$ & $2.46$ & $0.92$ & $0.74$ \\
    BigCodec-300M & $4.58$ & $3.24$ & $\mathbf{2.68}$ & $\mathbf{0.93}$ & $\mathbf{0.84}$ \\
    BigCodec-ll60k & $4.59$ & $3.24$ & $2.67$ & $\mathbf{0.93}$ & $\mathbf{0.84}$ \\
    small-enc & $4.61$ & $3.19$ & $2.60$ & $\mathbf{0.93}$ & $0.81$ \\
    \midrule
    w/o LSTM & $4.71$ & $3.16$ & $2.58$ & $0.92$ & $0.82$ \\
    \bottomrule
    \end{tabular}
    \label{tab:ablation}
\end{table}
\vspace{-2mm}
\subsection{Ablation studies}
\vspace{-1mm}
\label{subsection:ablation_studies}
Next, we conduct ablation studies.
Specifically, we examine five settings: (1) BigCodec with a smaller encoder and decoder (BigCodec-base), (2) BigCodec-300M, which scales up the model size to 300M, (3) BigCodec-ll60k, which trains BigCodec on the Libri-Light corpus~\cite{2020librilight} with 60k hours of data, (4) BigCodec with a smaller encoder (small-enc), as previous work suggests that the decoder dominates the reconstruction quality~\cite{wu2023audiodec}, and (5) BigCodec without the LSTM (w/o LSTM).
Note that, in the w/o LSTM setting, removing the two LSTM in BigCodec would nearly halve the number of parameters to $84$M, causing an unfair comparison.
To mitigate this, we increase the final channel size of the CNN in the encoder to $2048$, yielding a model with $142$M parameters.
The results are shown in Table~\ref{tab:ablation}.
First, BigCodec-base shows significantly lower performance than BigCodec, highlighting the effectiveness of scaling up the model size.
However, BigCodec-base still achieves performance comparable to the best baseline, TF-Codec, as shown in Table~\ref{tab:main_results}, despite the use of sophisticated autoregressive predictive coding by TF-Codec.
This demonstrates the effectiveness of BigCodec's architecture.
Second, BigCodec-300M achieves only comparable performance to BigCodec, showing no further gains from scaling up the model size, similar to the observation in BigVGAN~\cite{lee2022bigvgan}.
Third, BigCodec-ll60k also performs comparably to BigCodec, suggesting no clear advantage from scaling up the data size in codecs, although recent work suggests such benefits in generative models~\cite{ju2024ns3}.
Next, small-enc shows slightly lower performance than BigCodec, indicating that a sufficiently large encoder is also necessary to achieve optimal performance.
Finally, the w/o LSTM model underperforms BigCodec, underscoring the importance of modeling temporal dependency in codecs.

\vspace{-2mm}
\subsection{Efficiency analysis}
\vspace{-1mm}
Finally, we analyze the inference efficiency of BigCodec.
We approximate the real time factor (RTF) of BigCodec and BigCodec-base on an Intel(R) Core(TM) i9-7920X CPU @ 2.90GHz using the entire test set with $2620$ utterances.
A RTF greater than one indicates that the model can process data in real time.
The RTFs for both encoding and decoding in BigCodec are $1.1 \times$, while BigCodec-base achieves $3.1 \times$.
The RTF of BigCodec being only slightly above $1$ reflects the trade-off for its high performance, which is expected given the model’s complexity.
\vspace{-3mm}
\section{Conclusions}
\vspace{-1mm}
In this paper, we presented BigCodec, a neural speech codec that pushes the limits of low-bitrate codecs.
BigCodec achieves substantial advancements in low-bitrate speech coding by effectively scaling up model size and optimizing key components such as vector quantization and temporal dependency modeling.
Comprehensive objective and subjective evaluations on the in-domain dataset and unseen languages demonstrate that BigCodec, with a model size of $159$M parameters and a bitrate of $1.04$ kbps, significantly outperforms previous low-bitrate codecs.
Additionally, BigCodec achieves objective performance comparable to popular codecs operating at 4-6 times higher bitrates, and even delivers better subjective perceptual quality than the GT.
In the future we plan to generalize BigCodec to arbitrary audio data and explore further reductions in bitrate.
{\normalsize
\\\textbf{Acknowledgements:} 
We would like to thank Xue Jiang for providing the test samples of TF-Codec, as well as Youqiang Zheng and Weiping Tu for their assistance in evaluating SRCodec and SuperCodec.
We thank Yuki Saito for reviewing this work.
}

\bibliographystyle{IEEEtran}
\bibliography{mybib}

\begin{thebibliography}{10}
\providecommand{\url}[1]{#1}
\csname url@samestyle\endcsname
\providecommand{\newblock}{\relax}
\providecommand{\bibinfo}[2]{#2}
\providecommand{\BIBentrySTDinterwordspacing}{\spaceskip=0pt\relax}
\providecommand{\BIBentryALTinterwordstretchfactor}{4}
\providecommand{\BIBentryALTinterwordspacing}{\spaceskip=\fontdimen2\font plus
\BIBentryALTinterwordstretchfactor\fontdimen3\font minus \fontdimen4\font\relax}
\providecommand{\BIBforeignlanguage}[2]{{%
\expandafter\ifx\csname l@#1\endcsname\relax
\typeout{** WARNING: IEEEtran.bst: No hyphenation pattern has been}%
\typeout{** loaded for the language `#1'. Using the pattern for}%
\typeout{** the default language instead.}%
\else
\language=\csname l@#1\endcsname
\fi
#2}}
\providecommand{\BIBdecl}{\relax}
\BIBdecl

\bibitem{opus2012}
\BIBentryALTinterwordspacing
J.-M. Valin, K.~Vos, and T.~B. Terriberry, ``{Definition of the Opus Audio Codec},'' RFC 6716, Sep. 2012. [Online]. Available: \url{https://www.rfc-editor.org/info/rfc6716}
\BIBentrySTDinterwordspacing

\bibitem{evs2015}
M.~Dietz, M.~Multrus, V.~Eksler, V.~Malenovsky, E.~Norvell, H.~Pobloth, L.~Miao, Z.~Wang, L.~Laaksonen, A.~Vasilache \emph{et~al.}, ``Overview of the evs codec architecture,'' in \emph{Proc. ICASSP}.\hskip 1em plus 0.5em minus 0.4em\relax IEEE, 2015, pp. 5698--5702.

\bibitem{zeghidour2021soundstream}
N.~Zeghidour, A.~Luebs, A.~Omran, J.~Skoglund, and M.~Tagliasacchi, ``Soundstream: An end-to-end neural audio codec,'' \emph{IEEE/ACM Transactions on Audio, Speech, and Language Processing}, vol.~30, pp. 495--507, 2021.

\bibitem{jayashankar2022architecture}
T.~Jayashankar, T.~Koehler, K.~Kalgaonkar, Z.~Xiu, J.~Wu, J.~Lin, P.~Agrawal, and Q.~He, ``Architecture for variable bitrate neural speech codec with configurable computation complexity,'' in \emph{Proc. ICASSP}.\hskip 1em plus 0.5em minus 0.4em\relax IEEE, 2022, pp. 861--865.

\bibitem{defossez2022encodec}
A.~D{\'e}fossez, J.~Copet, G.~Synnaeve, and Y.~Adi, ``High fidelity neural audio compression,'' \emph{arXiv preprint arXiv:2210.13438}, 2022.

\bibitem{kumar2024descript}
R.~Kumar, P.~Seetharaman, A.~Luebs, I.~Kumar, and K.~Kumar, ``High-fidelity audio compression with improved rvqgan,'' in \emph{Proc. NeurIPS}, vol.~36, 2024.

\bibitem{denes1963statistics}
P.~B. Denes, ``On the statistics of spoken english,'' \emph{The Journal of the Acoustical Society of America}, vol.~35, no.~6, pp. 892--904, 1963.

\bibitem{van2017information}
S.~Van~Kuyk, W.~B. Kleijn, and R.~C. Hendriks, ``On the information rate of speech communication,'' in \emph{Proc. ICASSP}.\hskip 1em plus 0.5em minus 0.4em\relax IEEE, 2017, pp. 5625--5629.

\bibitem{tishby2000bottleneck}
N.~Tishby, F.~C. Pereira, and W.~Bialek, ``The information bottleneck method,'' \emph{arXiv preprint physics/0004057}, 2000.

\bibitem{kleijn2018wavenet}
W.~B. Kleijn, F.~S. Lim, A.~Luebs, J.~Skoglund, F.~Stimberg, Q.~Wang, and T.~C. Walters, ``Wavenet based low rate speech coding,'' in \emph{Proc. ICASSP}.\hskip 1em plus 0.5em minus 0.4em\relax IEEE, 2018, pp. 676--680.

\bibitem{siahkoohi2022ultra}
A.~Siahkoohi, M.~Chinen, T.~Denton, W.~B. Kleijn, and J.~Skoglund, ``Ultra-low-bitrate speech coding with pretrained transformers,'' \emph{arXiv preprint arXiv:2207.02262}, 2022.

\bibitem{zheng2024srcodec}
Y.~Zheng, W.~Tu, L.~Xiao, and X.~Xu, ``Srcodec: Split-residual vector quantization for neural speech codec,'' in \emph{Proc. ICASSP}.\hskip 1em plus 0.5em minus 0.4em\relax IEEE, 2024, pp. 451--455.

\bibitem{zheng2024supercodec}
------, ``Supercodec: A neural speech codec with selective back-projection network,'' in \emph{Proc. ICASSP}.\hskip 1em plus 0.5em minus 0.4em\relax IEEE, 2024, pp. 566--570.

\bibitem{jiang2023latent}
X.~Jiang, X.~Peng, H.~Xue, Y.~Zhang, and Y.~Lu, ``Latent-domain predictive neural speech coding,'' \emph{IEEE/ACM Transactions on Audio, Speech, and Language Processing}, vol.~31, pp. 2111--2123, 2023.

\bibitem{jenrungrot2023lmcodec}
T.~Jenrungrot, M.~Chinen, W.~B. Kleijn, J.~Skoglund, Z.~Borsos, N.~Zeghidour, and M.~Tagliasacchi, ``Lmcodec: A low bitrate speech codec with causal transformer models,'' in \emph{Proc. ICASSP}.\hskip 1em plus 0.5em minus 0.4em\relax IEEE, 2023, pp. 1--5.

\bibitem{brown2020gpt3}
T.~Brown, B.~Mann, N.~Ryder, M.~Subbiah, J.~D. Kaplan, P.~Dhariwal, A.~Neelakantan, P.~Shyam, G.~Sastry, A.~Askell \emph{et~al.}, ``Language models are few-shot learners,'' in \emph{Proc. NeurIPS}, vol.~33, 2020, pp. 1877--1901.

\bibitem{ramesh2021text2image}
A.~Ramesh, M.~Pavlov, G.~Goh, S.~Gray, C.~Voss, A.~Radford, M.~Chen, and I.~Sutskever, ``Zero-shot text-to-image generation,'' in \emph{Proc. ICML}.\hskip 1em plus 0.5em minus 0.4em\relax Pmlr, 2021, pp. 8821--8831.

\bibitem{lee2022bigvgan}
S.-g. Lee, W.~Ping, B.~Ginsburg, B.~Catanzaro, and S.~Yoon, ``Bigvgan: A universal neural vocoder with large-scale training,'' \emph{arXiv preprint arXiv:2206.04658}, 2022.

\bibitem{goodfellow2020generative}
I.~Goodfellow, J.~Pouget-Abadie, M.~Mirza, B.~Xu, D.~Warde-Farley, S.~Ozair, A.~Courville, and Y.~Bengio, ``Generative adversarial networks,'' \emph{Communications of the ACM}, vol.~63, no.~11, pp. 139--144, 2020.

\bibitem{van2017vqvae}
A.~van~den Oord, O.~Vinyals, and k.~kavukcuoglu, ``Neural discrete representation learning,'' in \emph{Proc. NeurIPS}, I.~Guyon, U.~V. Luxburg, S.~Bengio, H.~Wallach, R.~Fergus, S.~Vishwanathan, and R.~Garnett, Eds., vol.~30.\hskip 1em plus 0.5em minus 0.4em\relax Curran Associates, Inc., 2017.

\bibitem{panayotov2015librispeech}
V.~Panayotov, G.~Chen, D.~Povey, and S.~Khudanpur, ``Librispeech: an asr corpus based on public domain audio books,'' in \emph{Proc. ICASSP}.\hskip 1em plus 0.5em minus 0.4em\relax IEEE, 2015, pp. 5206--5210.

\bibitem{atal1982predictive}
B.~Atal, ``Predictive coding of speech at low bit rates,'' \emph{IEEE Transactions on Communications}, vol.~30, no.~4, pp. 600--614, 1982.

\bibitem{yang2024uniaudio15}
D.~Yang, H.~Guo, Y.~Wang, R.~Huang, X.~Li, X.~Tan, X.~Wu, and H.~Meng, ``Uniaudio 1.5: Large language model-driven audio codec is a few-shot audio task learner,'' \emph{arXiv preprint arXiv:2406.10056}, 2024.

\bibitem{ziyin2020snake}
L.~Ziyin, T.~Hartwig, and M.~Ueda, ``Neural networks fail to learn periodic functions and how to fix it,'' in \emph{Proc. NeurIPS}, vol.~33, 2020, pp. 1583--1594.

\bibitem{yu2021vector}
J.~Yu, X.~Li, J.~Y. Koh, H.~Zhang, R.~Pang, J.~Qin, A.~Ku, Y.~Xu, J.~Baldridge, and Y.~Wu, ``Vector-quantized image modeling with improved vqgan,'' \emph{arXiv preprint arXiv:2110.04627}, 2021.

\bibitem{kumar2019melgan}
K.~Kumar, R.~Kumar, T.~De~Boissiere, L.~Gestin, W.~Z. Teoh, J.~Sotelo, A.~De~Brebisson, Y.~Bengio, and A.~C. Courville, ``Melgan: Generative adversarial networks for conditional waveform synthesis,'' in \emph{Proc. NeurIPS}, vol.~32, 2019.

\bibitem{kong2020hifi}
J.~Kong, J.~Kim, and J.~Bae, ``Hifi-gan: Generative adversarial networks for efficient and high fidelity speech synthesis,'' in \emph{Proc. NeurIPS}, vol.~33, 2020, pp. 17\,022--17\,033.

\bibitem{jang21univnet}
W.~Jang, D.~Lim, J.~Yoon, B.~Kim, and J.~Kim, ``{UnivNet: A Neural Vocoder with Multi-Resolution Spectrogram Discriminators for High-Fidelity Waveform Generation},'' in \emph{Proc. Interspeech}, 2021, pp. 2207--2211.

\bibitem{yamamoto2020parallel}
R.~Yamamoto, E.~Song, and J.-M. Kim, ``Parallel wavegan: A fast waveform generation model based on generative adversarial networks with multi-resolution spectrogram,'' in \emph{Proc. ICASSP}.\hskip 1em plus 0.5em minus 0.4em\relax IEEE, 2020, pp. 6199--6203.

\bibitem{mao2017lsgan}
X.~Mao, Q.~Li, H.~Xie, R.~Y. Lau, Z.~Wang, and S.~Paul~Smolley, ``Least squares generative adversarial networks,'' in \emph{Proc. ICCV}, 2017, pp. 2794--2802.

\bibitem{bengio2013straight_through}
Y.~Bengio, N.~L{\'e}onard, and A.~Courville, ``Estimating or propagating gradients through stochastic neurons for conditional computation,'' \emph{arXiv preprint arXiv:1308.3432}, 2013.

\bibitem{loshchilov2018adamw}
I.~Loshchilov and F.~Hutter, ``Decoupled weight decay regularization,'' in \emph{Proc. ICLR}, 2018.

\bibitem{reddy2021icassp}
C.~K. Reddy, H.~Dubey, V.~Gopal, R.~Cutler, S.~Braun, H.~Gamper, R.~Aichner, and S.~Srinivasan, ``Icassp 2021 deep noise suppression challenge,'' in \emph{Proc. ICASSP}.\hskip 1em plus 0.5em minus 0.4em\relax IEEE, 2021, pp. 6623--6627.

\bibitem{pratap20mls}
V.~Pratap, Q.~Xu, A.~Sriram, G.~Synnaeve, and R.~Collobert, ``{MLS: A Large-Scale Multilingual Dataset for Speech Research},'' in \emph{Proc. Interspeech}, 2020, pp. 2757--2761.

\bibitem{kim2019audiocaps}
C.~D. Kim, B.~Kim, H.~Lee, and G.~Kim, ``Audiocaps: Generating captions for audios in the wild,'' in \emph{Proc. NAACL}, 2019, pp. 119--132.

\bibitem{rix2001pesq}
A.~W. Rix, J.~G. Beerends, M.~P. Hollier, and A.~P. Hekstra, ``Perceptual evaluation of speech quality (pesq)-a new method for speech quality assessment of telephone networks and codecs,'' in \emph{Proc. ICASSP}, vol.~2.\hskip 1em plus 0.5em minus 0.4em\relax IEEE, 2001, pp. 749--752.

\bibitem{taal2011stoi}
C.~H. Taal, R.~C. Hendriks, R.~Heusdens, and J.~Jensen, ``An algorithm for intelligibility prediction of time--frequency weighted noisy speech,'' \emph{IEEE Transactions on audio, speech, and language processing}, vol.~19, no.~7, pp. 2125--2136, 2011.

\bibitem{chen2022wavlm}
S.~Chen, C.~Wang, Z.~Chen, Y.~Wu, S.~Liu, Z.~Chen, J.~Li, N.~Kanda, T.~Yoshioka, X.~Xiao \emph{et~al.}, ``Wavlm: Large-scale self-supervised pre-training for full stack speech processing,'' \emph{IEEE Journal of Selected Topics in Signal Processing}, vol.~16, no.~6, pp. 1505--1518, 2022.

\bibitem{series2014mushra}
B.~Series, ``Method for the subjective assessment of intermediate quality level of audio systems,'' \emph{International Telecommunication Union Radiocommunication Assembly}, vol.~2, 2014.

\bibitem{schoeffler2018webmushra}
M.~Schoeffler, S.~Bartoschek, F.-R. Stoter, M.~Roess, S.~Westphal, B.~Edler, and J.~Herre, ``webmushra-a comprehensive framework for web-based listening tests,'' \emph{Journal of Open Research Software}, vol.~6, no.~7, 2018.

\bibitem{cicchetti1994guidelines}
D.~V. Cicchetti, ``Guidelines, criteria, and rules of thumb for evaluating normed and standardized assessment instruments in psychology.'' \emph{Psychological assessment}, vol.~6, no.~4, p. 284, 1994.

\bibitem{2020librilight}
J.~{Kahn}, M.~{Rivière}, W.~{Zheng}, E.~{Kharitonov}, Q.~{Xu}, P.~E. {Mazaré}, J.~{Karadayi}, V.~{Liptchinsky}, R.~{Collobert}, C.~{Fuegen}, T.~{Likhomanenko}, G.~{Synnaeve}, A.~{Joulin}, A.~{Mohamed}, and E.~{Dupoux}, ``Libri-light: A benchmark for asr with limited or no supervision,'' in \emph{Proc. ICASSP}, 2020, pp. 7669--7673, \url{https://github.com/facebookresearch/libri-light}.

\bibitem{wu2023audiodec}
Y.-C. Wu, I.~D. Gebru, D.~Markovi{\'c}, and A.~Richard, ``Audiodec: An open-source streaming high-fidelity neural audio codec,'' in \emph{Proc. ICASSP}.\hskip 1em plus 0.5em minus 0.4em\relax IEEE, 2023, pp. 1--5.

\bibitem{ju2024ns3}
Z.~Ju, Y.~Wang, K.~Shen, X.~Tan, D.~Xin, D.~Yang, Y.~Liu, Y.~Leng, K.~Song, S.~Tang \emph{et~al.}, ``Naturalspeech 3: Zero-shot speech synthesis with factorized codec and diffusion models,'' \emph{arXiv preprint arXiv:2403.03100}, 2024.

\end{thebibliography}

\end{document}